# Fast diffusion of water nanodroplets on graphene

Ming Ma[1,2], Gabriele Tocci[1], Angelos Michaelides[1,2,*], and Gabriel Aeppli[1,3,4]

[1]London Centre for Nanotechnology, 17–19 Gordon Street, London WC1H 0AH, UK,

[2]Department of Chemistry, University College London, London WC1H 0AJ, UK

[3]ETH Zürich CH-8093, EPF Lausanne CH-1015 and Paul Scherrer Institute, Villigen CH-5232, Switzerland

[4]Bio Nano Consulting, The Gridiron Building, One Pancras Square, London N1C 4AG, UK

Diffusion across surfaces generally involves motion on a vibrating but otherwise stationary substrate. Here, using molecular dynamics, we show that a layered material such as graphene opens up a new mechanism for surface diffusion whereby adsorbates are carried by propagating ripples via a motion similar to surfing. For water nanodroplets, we demonstrate that the mechanism leads to exceedingly fast diffusion that is 2-3 orders of magnitude faster than the self-diffusion of water molecules in liquid water. We also reveal the underlying principles that regulate this new mechanism for diffusion and show how it also applies to adsorbates other than water, thus opening up the prospect of achieving fast and controllable motion of adsorbates across material surfaces more generally.

* email: angelos.michaelides@ucl.ac.uk

The motion of atoms, molecules and clusters across the surfaces of materials is of critical importance to an endless list of phenomena. Consider, for example, the diffusion of molecules across the surfaces of catalyst particles in search of reaction partners or active sites. Likewise, consider how the elementary building blocks of a growing crystal must diffuse over the faces of that crystal to find the correct lattice sites to attach to. Similarly, at the heart of chemical vapor deposition and many self-assembly and filtration processes is the diffusion of particles across surfaces. Generally when thinking about the atomic-scale motion of adsorbates on surfaces, one envisages a scenario where the adsorbates move (or hop) on the surface while the vibrating but otherwise stationary substrates provide the thermal energy for the motion (see e.g. Refs.[1-3]).

A whole arsenal of experimental techniques has been developed to probe the atomic-scale details of adsorbate diffusion[1-3]. Prominent methodologies include field ion microscopy, scanning probes, and helium atom scattering. Generally these approaches have been applied to ultra-clean atomically flat surfaces prepared in ultra high vacuum. Although a number of novel diffusion mechanisms have been proposed (see e.g. Refs.[2,4]), for the most part experiments have supported the notion that diffusion involves the rather straightforward motion of the adsorbate from one lattice site to the next. However, most well-defined diffusion studies have been on the surfaces of "traditional" three-dimensional crystalline materials including metals, semi-conductors and oxides[1-3]. Relatively little work has been done to explore diffusion on thin layered solids[5,6]. This is true despite the surge of interest in the properties of layered materials and their interactions with adsorbates (see e.g. Refs.[6-9]).

Of all adsorption systems involving layered materials, the one with the greatest technological relevance and attracting most attention at present is water on graphene[8,10-14]. It is being explored because of its direct impact on graphene-based devices, such as water filtration membranes[15], and also as a model system to understand the interaction between water and carbonaceous surfaces. Recent water-graphene studies (either on free standing or supported graphene) have focused on wetting[10,11,16], structure[11,14] reactivity[8], and the motion and manipulation of (saline) water nanodroplets[12]. Concerning diffusion there has been a large body of excellent work examining water on flat surfaces or in various confining geometries, including the widely studied processes of water diffusion and transport across and within carbon nanotubes (see e.g. Refs.[11,16-23]). However, understanding of water diffusion on graphene and layered materials in general is in its infancy. In particular it is not known how the large out-of-plane displacements exhibited by single layers of these materials might impact adsorbate motion[24-26].

Here we report results from an extensive and carefully validated set of force field-based molecular dynamics (MD) simulations for the diffusion of water nanodroplets on graphene as well as two distinctly different hydrocarbons. The simulations reveal an interesting and potentially important new physical mechanism for diffusion on the nanoscale involving coupling of the adsorbate to thermally excited, propagating ripples on graphene in a manner that resembles surfing. Through being swept along by the motion of the substrate water nanodroplets diffuse rapidly, with the smallest droplets diffusing at least an order of magnitude faster than reported previously for other adsorbates on related systems[27-32]. We explore and discuss the general principles underlying this novel mechanism for diffusion, specifically focusing on the roles of the ripple amplitude and the interaction strength between adsorbates and graphene sheets. Indeed we show that the principles identified explain some rather counterintuitive results, such as simulations that show faster water droplet diffusion on more corrugated graphene sheets and the fact that water droplets diffuse more rapidly than more weakly bound adsorbates such as $C_{60}$. We also provide evidence that the mechanism applies to hydrocarbon motion and discuss various experiments that could support our predictions as well as practical ways to exploit the phenomena observed.

**Ripples in graphene**

We begin with some brief background on clean graphene. A free-standing or suspended sheet of graphene is rippled[24-26,33,34], where the ripples are thermally and/or mechanically induced. Thermally activated ripples are caused by anharmonic coupling between the bending and stretching modes of graphene and have been rationalized on the basis of flexible membrane theories and atomistic simulations[24-26]. In particular the root mean square of the height of the ripples, $\bar{h}$, scales with the size of the graphene sheet being considered, $L_0$, as $\bar{h} = \alpha L_0^\eta$ with $0 < \eta < 1$ and $\alpha$ is a constant. Such a power law relation between $\bar{h}$ and $L_0$ implies that the ripples do not exhibit a characteristic wavelength below $L_0$. With the typical force fields used for atomistic simulations of graphene, the amplitude of individual ripples is on the order of a few Angstroms (Refs [24-26] and see also Section 2 in the supplementary information[35] (SI)). The lifetime of individual ripples is comparable to that of the out-of-plane acoustic (ZA) phonons with similar spatial scale, which is usually on the order of tens of ps to a few ns[36]. In practice, however, the ripples observed so far in experiments have amplitudes in the nanometer regime and well-defined separations between their crests[33,34,37-39]. The apparent discrepancy between theory and experiment most likely arises because of strain introduced at the edges and contact points of graphene by the supports or clamps that will inevitably be present in experiments[33]. Indeed very small compressive or shear strains induce significant buckling, as seen before[33] and shown in our simulations below. Recent scanning tunneling microscopy (STM) data show that the ripples in free-standing graphene are dynamic[34], although in the vicinity of the support the mobility is reduced which helps to explain why an earlier STM study saw static ripples[38].

We have carried out simulations for a range of unit cells with several carbon-carbon interaction potentials and with and without applied lateral strain. We find that the amplitude and the characteristic separations (if any) of the ripples depend sensitively on the precise set-up used. Small amplitude ripples such as those obtained in some previous simulation studies[24-26] or larger amplitude ripples, as characterized by STM[34,38], can both be readily produced. The main difference between the two types of ripples is their spatial coherence over the lifetime of the simulations: small amplitude ripples are short-lived whereas those with larger amplitudes are much longer-lived.

The larger amplitude ripples persist for the entire duration of our simulations (Fig. S1[35]) and exhibit soliton-like behavior, which is again consistent with some previous MD studies[40,41]. The transition between short and long-lived ripples is in quantitative agreement with that predicted by theory and observed in experiments (Fig. S1[35]).

**MD simulations for water nanodroplets on graphene**

Here we report and discuss results for water droplet diffusion on free-standing graphene sheets hosting both types of ripples. Most droplets were between 2 to 6 nm in diameter, although results on larger droplets are presented in Section 12 in the SI[35]. Such droplets have sizes approaching those observed in recent scanning probe experiments under ambient conditions[13,42], yet they are small enough to enable long (tens to hundreds of nanoseconds) MD trajectories from which accurate dynamical information can be obtained. Details of MD simulations can be found in the Methods section. Figure 1 shows snapshots of two intermediate sized nanodroplets diffusing on two differently rippled graphene surfaces. Movies of the diffusion processes are also provided in Section 16 in the SI. From Fig. 1 it can be inferred that the droplets inhabit the valleys of the graphene ripples (blue regions in Fig. 1) and as the ripples of graphene move so too do the water droplets in a mechanism reminiscent of surfing. Examination of the water droplets reveals that they are liquid with a higher density of water molecules in the immediate vicinity of the graphene (see Section 7 in the SI[35]). This is in line with previous *ab initio* and force field studies of liquid water on graphene[11,21,43]. By calculating the commensurability between the structure of the water molecules in the first layer and that of the underlying graphene we find no apparent registry between them. This is consistent with the potential energy surface for a water monomer on graphene being exceptionally flat, as found in our own force field studies and previous *ab initio* calculations (see e.g. [16]). The water droplet has a fairly small influence on the structure of the ripples: carbon atoms directly beneath the droplets are generally displaced by <0.02 nm from where they would be expected to be in the absence of the droplet.

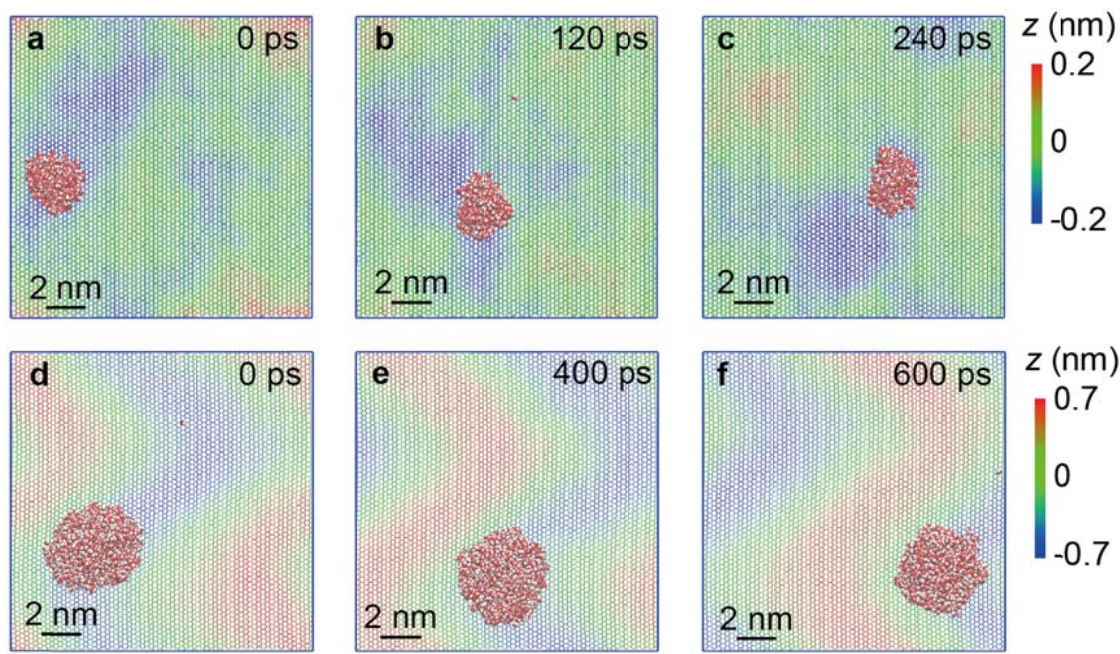

**Figure 1 | Water droplet diffusion on graphene**. Snapshots of a water nanodroplet diffusing on graphene with small (*ca*. 0.2 nm) amplitude ripples (top panels, **a-c**) and large (*ca*. 0.7 nm) amplitude ripples (bottom panels, **d-f**). The carbon atoms of the graphene are colour-coded according to their height, thus indicating that the water droplets reside within the valleys of the ripples (blue regions). In the snapshots shown the ripples advance from left to right carrying the water droplets with them, similar to how a surfer rides a wave. The particular droplets shown have diameters of about 2 nm (top row, ~205 water molecules) and 4.4 nm (bottom row, ~750 water molecules). The blue box around each panel indicates the periodically repeated unit cell, which has lateral dimensions of 15×15 nm.

We find that the droplets diffuse very rapidly and that the speed of the diffusion is strongly size-dependent. This can be seen in Fig. 2a where the diffusion coefficient ($D$) of the droplets is plotted as a function of droplet size for a graphene sheet with small amplitude ripples. Results for the large amplitude ripples are given in Fig. S8. Over the size regime shown $D$ varies from $2.0\times10^{-3} \pm 3\times10^{-5}$ cm$^2$/s for the largest droplet (about 6 nm in diameter with about 1,374 water molecules) to $8.6\times10^{-3} \pm 2\times10^{-5}$ cm$^2$/s for the smallest droplet (about 2 nm in diameter with about 105 water molecules). As shown by the dashed line in Fig. 2a the dependence of $D$ on the number of water molecules $N$ within the droplet can be reasonably described by a simple model based on the surface-of-contact concept as $D \sim \alpha N^{-2/3}$, where $\alpha$ is a constant (see Section 8 in the SI[35] for a discussion on how this emerges). Besides the size dependence of $D$ on $N$, we also note that even after accounting for the sensitivity of $D$ to the specific computational set-up used[35], the diffusion coefficients observed are exceptionally large and much greater than diffusion coefficients reported before for related systems[27-32]. We show this, for example, in Fig. 2b where we compare our results for water droplets with results from previous simulations of diffusion on graphene. Included in this comparison are $C_{60}$[27], a graphitic flake[28], and several metal clusters[29-32]. In all cases the water droplets diffuse at least one order of magnitude faster than the other adsorbates and for most of them several orders of magnitude faster. The comparison in Fig. 2b is limited to simulations and to diffusion on graphene. Comparison with other systems further underscores the fast diffusion coefficients predicted here. For example, predicted diffusion coefficients of water through carbon nanotubes (which host an inherently different diffusion process that relies on low friction between water and the nanotube) are about 40 times lower than what we find here[44]. Likewise the diffusion of the droplets seen here is about 2-3 orders of magnitude faster than the self-diffusion of water molecules in liquid water [45].

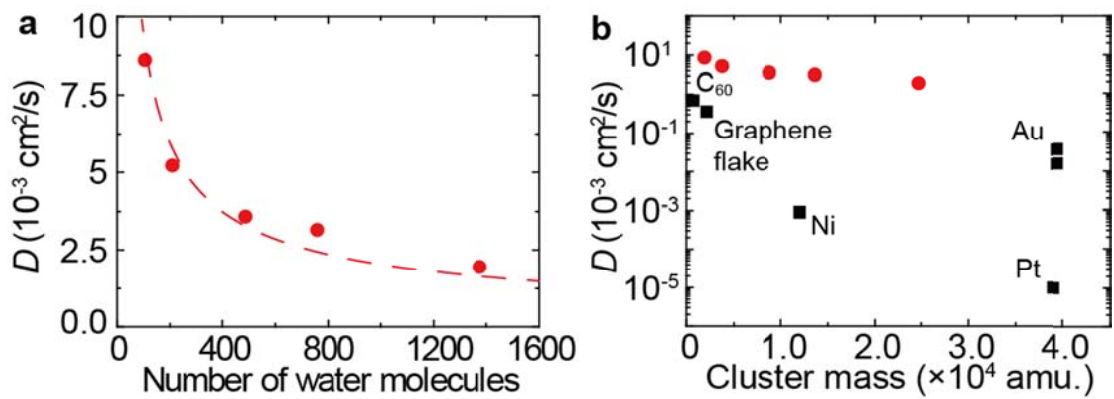

**Figure 2 | Diffusion coefficients $D$ for different sized water nanodroplets and comparison between water nanodroplets and other adsorbates**. **a**, Diffusion coefficients of water nanodroplets on graphene with small amplitude ripples (~0.2 nm)

as a function of the number of water molecules within the nanodroplets. The dashed line is a one-parameter fit of the form $D \sim \alpha N^{-2/3}$ based on the surface-of-contact concept[35]. **b**, Comparison between water nanodroplet diffusion computed in this work (red dots) and the diffusion coefficients for other adsorbed clusters on graphene shown on a semi-log plot. Previously published simulation results are from: $C_{60}$[27], graphene flake[28] (comprised of 178 C atoms), Ni[29], Au[30,31], and Pt[32]. The statistical error bar on each water data point is smaller that the actual size of the points displayed.

**Surfing mechanism for fast diffusion**

To understand the rapid diffusion of the water droplets in more detail, we calculated the diffusion coefficient $D$ as a function of the amplitude, $A$, of the ripples in the graphene sheet. We varied the amplitude by applying in-plane strain to graphene from −0.8% to 0.5%. This strain range is accessible experimentally[33,46] and yields ripples with amplitudes between 0.1 to 0.7 nm. As shown in Fig. 3a, for the ripple amplitudes considered, $D$ increases with $A$. Thus water droplets diffuse more rapidly on the more corrugated surfaces. We find that this counterintuitive result arises from the greater coupling of the water droplets to the more corrugated surfaces. To quantify the coupling we consider a probability density function, $PDF(z)$, of the average height ($z$) of the carbon atoms beneath the water droplet with respect to the average height of the graphene sheet. As shown in Fig. 3b, as the amplitude of the ripples increases, $PDF(z)$ peaks at increasingly negative values of $z$, revealing an increased preference of the droplets for the valleys. Taking this one step further we plot in Fig. 3c how the droplet diffusion coefficient depends on the probability $P(z < 0) = \int_{-\infty}^{0} PDF(z)dz$, for water droplets to inhabit low-lying regions of grapheme. Values of $P(z < 0)$ close to unity indicate that water droplets are almost always found in valleys, whereas a value of 0.5 would indicate no preference for either the low-lying or high-lying regions of graphene. From Fig. 3c it is evident that the diffusion coefficient increases as $P(z < 0)$ increases. Note that for large amplitude ripples ($A$~0.66 nm), the water droplets are almost always carried by the ripples ($P(z < 0) \sim 1.00$) but that even for very small amplitude ripples ($A \sim 0.08$ nm), the water droplets still show a tendency to couple with the ripples ($P(z < 0) \sim 0.70$). This strong correlation clearly shows that it is the coupling with the ripples that enhances the diffusion of water droplets. We also note that this coupling with the ripples occurs irrespective of the lifetime of the ripples. The dashed vertical lines in Figs. 3a and 3c mark the transition between long-lived coherent ripples and short-lived incoherent ripples (see Fig. S1d in the SI[35] for more details). No obvious change in droplet coupling or diffusion mechanism is observed upon moving between these two regimes. Such dependence of $D$ on $A$ holds for all water droplets considered as shown in Section 9 in the SI[35].

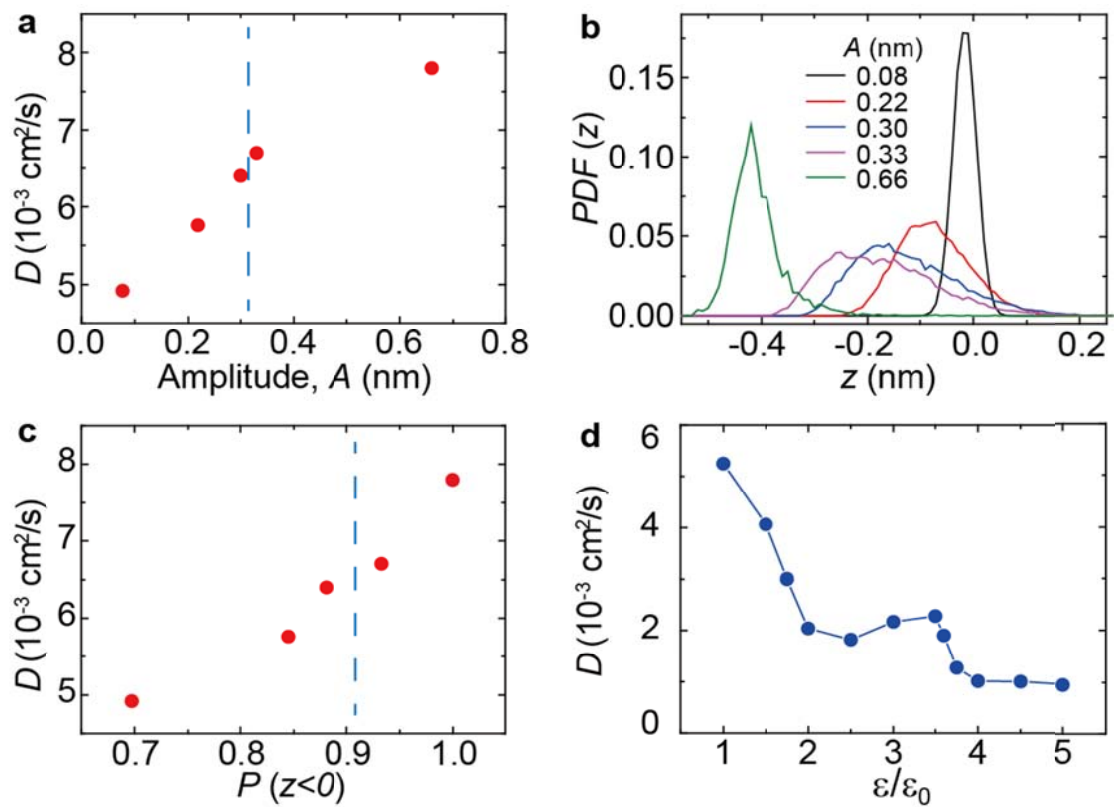

**Figure 3 | Connection between diffusion and the amplitude of the ripples of graphene and the adsorbate/substrate interaction strength**. **a**, Diffusion coefficient $D$ as a function of ripple amplitude $A$ for water droplets composed of ~ 205 water molecules. The points at $A = 0.22$ and $0.66$ nm correspond to the "small" and "large" amplitude ripples discussed throughout the text. **b**, Probability density

functions (PDF) of the average $z$ height of the carbon atoms below the water droplet with respect to the average height of the graphene sheet for the different graphene sheets reported in a. **c**, Diffusion coefficient $D$ as a function of the probability, $P(z < 0)$, for water droplets to be in the ripple valley regions. $P(z < 0)$ is given by $\int_{-\infty}^{0} PDF(z)dz$. **d**, Diffusion coefficient $D$ versus the $C_{60}$/graphene interaction strength ($\varepsilon/\varepsilon_0$) showing that for the diffusion of $C_{60}$ molecules on graphene a small region exists where $D$ increases as the interaction strength with the surface increases. $\varepsilon_0$ is simply a Lennard-Jones well depth taken from reference[27]. The solid line is merely a guide to the eye. Error bars are not shown in panels a, b, and d as the relative errors are all smaller than 0.3%. The dashed vertical lines in panels a and c mark the transition between long-lived coherent ripples and short-lived incoherent ripples.

As the motion of the ripples dictates the diffusion of the droplets, the ripple-droplet coupling is crucial to the surfing mechanism which requires for the adsorbate to have a preference for one particular region of the surface over others, so that it can be swept along by the movement of the ripples. Water, of course, exhibits such a preference and we expect many other adsorbates will too. For example, we show in Section 11 of the SI that dodecane droplets exploit the same surfing mechanism[35]. In contrast, if the adsorbate does not have a clear preference for one sign of graphene curvature, it will not surf but rather it will bob up and down on the surface like a buoy. We show an example of this contrasting behavior in Fig. 3d where we report results for the diffusion of $C_{60}$ on a graphene sheet with small amplitude ripples. A $C_{60}$ molecule was selected because our initial calculations indicated that it binds more weakly to the substrate than the water droplets and has a much smaller preference for the valleys than the water droplets. The diffusion coefficient of $C_{60}$ assuming a standard $C_{60}$/graphene interaction strength[27] is about 40% smaller than that of our smallest ~105 water molecule cluster despite the water cluster having a *ca*. 170 % greater mass, *c.f.* $D(C_{60}) = 5.2\times10^{-3} \pm 2 \times10^{-5}$ cm$^2$/s and $D(H_2O) = 8.6\times10^{-3} \pm 2 \times10^{-5}$ cm$^2$/s. We thus arrive at the counterintuitive result that the adsorbate that binds less strongly to the surface travels more slowly. The slower diffusion arises from less effective coupling with the surface: $P(z < 0) = 0.66$ for $C_{60}$ compared to $P(z < 0) = 0.85$ for the water droplet. Thus $C_{60}$ does not get swept along by the ripples to as great an extent as the water droplet but rather spends more time bobbing up and down on the surface. Extending this analysis further we explored the sensitivity of the $C_{60}$ diffusion coefficient to the strength of the interaction with the surface. This was done in our simulations by altering the Lennard-Jones well depth. We find in general that the diffusion coefficient decreases as the adsorbate-substrate interaction increases. However, as shown in Fig. 3d, we have identified a regime where an *increase* in the interaction strength leads to an *increase* in the rate of $C_{60}$ diffusion. This non-monotonic behavior of $D$ with respect to the $C_{60}$/graphene interaction strength indicates a non-trivial balance between the enhancement in $D$ induced by the coupling with the phonons and the reduction in $D$ induced by the strong interaction with the surface.

**Comparison with diffusion on other surfaces**

The sensitivity of the mechanism to the amplitude of the ripples is very important as it explains why this mechanism has not been observed on the surfaces of conventional crystalline materials. The amplitude of the out-of-plane displacements for travelling surface waves on the surfaces of more traditional materials, e.g. three-dimensional metals and oxides, is substantially smaller than the amplitude of the ripples on graphene[47]. Likewise for carbon nanotubes the amplitude of the radial breathing mode is significantly smaller than the amplitude of the ripples in graphene. Interestingly Král and co-workers have recently predicted that water nanodroplets outside of carbon nanotubes can be transported along the flexural mode at speeds of up to a few tens of m/s if the flexural mode is externally excited[23]. Similarly we expect that external excitation of the ripples will facilitate adsorbate motion on graphene. However, here we see that because of the much larger out-of-plane displacements in this two-dimensional "solid" thermal excitations are sufficient to facilitate rapid diffusion (see Section 15 in the SI). It is plausible that similar behavior will be seen on other thin layered materials[6] and for diffusion within porous crystalline materials[20] where there are also large amplitude displacements associated with flexure.

**Outlook**

Direct experimental verification of our microscopic simulations will require experiments which access the relevant length and time scales. A starting point would be the classic surface science tools of scanning probe microscopy and helium atom scattering; for the former, the location of water droplets formed by condensation onto supported graphene flakes could be found after freezing to low temperatures, and the preference for ripple valleys could be confirmed. Indeed under ultra-high vacuum conditions water monomers and sub-nanometer water clusters have been observed on metal surfaces[17] and even under ambient conditions nanodroplets in the ~10-100 nm size regime have been imaged on various substrates[42]. Similarly the dynamics of ripples in freestanding graphene has recently been observed experimentally on the atomic scale using STM at room temperature[34]. Helium atom spin echo spectroscopy can be used to observe the motion of both carbon and water, testing our dynamical results. These various ultra-high vacuum studies would most likely be performed at low temperatures to limit evaporation. In Section 13 in the SI[35] we show that the same basic mechanism also applies at a lower temperature of 200 K and does not require the droplets to be liquid. There are other experiments for which the conditions are closer to those characterizing applications. One example is surface acoustic wave spectroscopy which might be performed either using Brillouin (light) scattering on free-standing graphene or on integrated device structures built on SiC. The relative humidity could be changed and the damping and sound velocities could be monitored (see e.g. Ref.[48]); if the experiments bear out the dramatic effects which we predict, the device structures might eventually be used as humidity sensors. Other experiments to perform include nuclear magnetic resonance, X-ray and neutron scattering on ensembles and even individual (in the case of high brilliance X-ray sources) graphene flakes; these techniques give a variety of momentum and time-dependent correlation functions, with further information available from isotopic substitution studies.

For all experiments, quasi-free standing graphene on flat substrates such as Pt and SiC or pillars would be prime candidates to explore because of the weak coupling of graphene with these substrates[7]. A pertinent issue in connecting our results with experiment is the amplitude and the wavelength of the ripples formed. These are physical characteristics of graphene that will depend on the precise system being explored. As we show in Fig. 3 our simulations predict an interesting and strong sensitivity of the results to the amplitude of the ripples. In contrast, water droplet diffusion is less likely to be strongly dependent on the wavelength of the ripples[35]. However if the droplets are significantly larger than the distance between ripple crests such that a single droplet straddles several ripples we may move out of the surfing regime. Defects will invariably be present in any experimentally realizable system and are certain to influence the dynamics, especially if water is pinned to the defects[42] or reacts with them as was seen recently for water on the strongly coupled graphene/Ru system[8]. If, however, defects serve merely to alter the dynamics of the ripples the physical picture arrived at here is still applicable for water droplet diffusion, as we discuss in Section 10 in the SI[35].

In conclusion, the fast diffusion of water nanodroplets on graphene is explained through a novel diffusion mechanism. Unlike diffusion on solid surfaces, where the substrate merely serves as a thermal bath, here we find that adsorbate and substrate motion is strongly coupled. Indeed we have identified a mechanism wherein the water droplets co-move with ripples of the underlying substrate. We are not aware of any previous studies reporting this phenomenon and some of the key features relevant to whether an adsorbate can diffuse rapidly via this mechanism have been discussed, including the amplitude of the ripples, the mass of the adsorbate and its ability to couple to the waves. Further, we have demonstrated that the novel mechanism extends beyond water to, at least, hydrocarbons and that the balance of effects can lead to seemingly counterintuitive results with reduced water diffusion on a less rippled surface or a more weakly bonded hydrocarbon that diffuses less rapidly than water. These examples are just a small subset of the outcomes of strong coupling of adsorbates to out-of-plane ripples in layered materials, which represents a new paradigm for surface diffusion. The principles and factors identified here as being critical to rapid diffusion, together with the possibility to tune the dispersion relation of graphene, graphene nanoribbons and other thin layered materials[33] as well as the height of the ripples[33] means that fast and controllable diffusion could be within reach. The ability to coat surfaces with graphene and other lamellar substances as well as the possibility of dissolving molecules, including proteins and DNA, in the water droplets may also enable highly controlled transport of chemicals across a broad range of materials. The enhanced transport *across* the surface could of course be the factor that makes the holes in such materials particularly accessible for transport *through* the surfaces, which is essential, for example in the case of the proposed nucleic acid readout devices[9].

**Acknowledgements**

M. M. was supported by the European Research Council (ERC) and Bio Nano Consulting. A.M. was also supported by the ERC and the Royal Society through a Royal Society Wolfson Research Merit Award. We are grateful for computer time to UCL Research Computing, the London Centre for Nanotechnology, and the UK's national high performance computing service HECToR (from which access was obtained via the UK's Material Chemistry Consortium, EP/F067496).

**Author contributions**

A.M., G.A. and M.M. proposed and designed the project. M.M. performed the force field MD simulations and data analysis. A.M. participated in data analysis. G.T. performed AIMD simulations. A.M. and M.M. wrote the manuscript. All authors participated in manuscript preparation.

**References**

1  Jardine, A.P., Hedgeland, H., Alexandrowicz, G., Allison, W. & Ellis, J. Helium-3 spin-echo: Principles and application to dynamics at surfaces. *Prog. Surf. Sci.* **84**, 323-379 (2009).
2  Antczak, G. & Ehrlich, G. Jump processes in surface diffusion. *Surf. Sci. Rep.*

**62**, 39-61 (2007).

3 Ala-Nissila, T., Ferrando, R. & Ying, S.C. Collective and single particle diffusion on surfaces. *Adv. Phys.* **51**, 949-1078 (2002).

4 Kellogg, G.L. & Feibelman, P.J. Surface self-diffusion on Pt(001) by an atomic exchange mechanism. *Phys. Rev. Lett.* **64**, 3143-3146 (1990).

5 Knudsen, J. *et al.* Clusters binding to the graphene moire on Ir(111): X-ray photoemission compared to density functional calculations. *Phys. Rev. B* **85**, 035407 (2012).

6 Geim, A.K. & Grigorieva, I.V. Van der Waals heterostructures. *Nature* **499**, 419-425 (2013).

7 Wintterlin, J. & Bocquet, M.-L. Graphene on metal surfaces. *Surf. Sci.* **603**, 1841-1852 (2009).

8 Feng, X.F., Maier, S. & Salmeron, M. Water splits epitaxial graphene and intercalates. *J. Am. Chem. Soc.* **134**, 5662-5668 (2012).

9 Garaj, S. *et al.* Graphene as a subnanometre trans-electrode membrane. *Nature* **467**, 190-U173 (2010).

10 Li, Z. *et al.* Effect of airborne contaminants on the wettability of supported graphene and graphite. *Nature Mater.* **12**, 925 - 931 (2013).

11 Cicero, G., Grossman, J.C., Schwegler, E., Gygi, F. & Galli, G. Water confined in nanotubes and between graphene sheets: A first principle study. *J. Am. Chem. Soc.* **130**, 1871-1878 (2008).

12 Yin, J. *et al.* Generating electricity by moving a droplet of ionic liquid along graphene. *Nature Nanotech.* **9**, 378-383 (2014).

13 Cheng, M. *et al.* A Route toward Digital Manipulation of Water Nanodroplets on Surfaces. *Acs Nano* **8**, 3955-3960 (2014).

14 Algara-Siller, G. *et al.* Square ice in graphene nanocapillaries. *Nature* **519**, 443-+ (2015).

15 Nair, R.R., Wu, H.A., Jayaram, P.N., Grigorieva, I.V. & Geim, A.K. Unimpeded permeation of water through helium-leak-tight graphene-based membranes. *Science* **335**, 442-444 (2012).

16 Ma, J. *et al.* Adsorption and diffusion of water on graphene from first principles. *Phys. Rev. B* **84**, 033402 (2011).

17 Mitsui, T., Rose, M.K., Fomin, E., Ogletree, D.F. & Salmeron, M. Water Diffusion and Clustering on Pd(111). *Science* **297**, 1850-1852 (2002).

18 Ranea, V.A. *et al.* Water dimer diffusion on Pd{111} assisted by an H-bond donor-acceptor tunneling exchange. *Phys. Rev. Lett.* **92**, 136104 (2004).

19 Carrasco, J., Hodgson, A. & Michaelides, A. A molecular perspective of water at metal interfaces. *Nature Mater.* **11**, 667-674 (2012).

20 Smit, B. & Maesen, T.L.M. Molecular simulations of zeolites: Adsorption, diffusion, and shape selectivity. *Chem. Rev.* **108**, 4125-4184 (2008).

21 Park, J.H. & Aluru, N.R. Ordering-induced fast diffusion of nanoscale water film on graphene. *J. Phys. Chem. C* **114**, 2595-2599 (2010).

22 Mittal, J., Truskett, T.M., Errington, J.R. & Hummer, G. Layering and position-dependent diffusive dynamics of confined fluids. *Phys. Rev. Lett.* **100**,

145901 (2008).

23  Russell, J.T., Wang, B.Y. & Kral, P. Nanodroplet Transport on Vibrated Nanotubes. *J. Phys. Chem. Lett.* **3**, 353-357 (2012).

24  Fasolino, A., Los, J.H. & Katsnelson, M.I. Intrinsic ripples in graphene. *Nature Mater.* **6**, 858-861 (2007).

25  Los, J.H., Katsnelson, M.I., Yazyev, O.V., Zakharchenko, K.V. & Fasolino, A. Scaling properties of flexible membranes from atomistic simulations: Application to graphene. *Phys. Rev. B* **80**, 121405(R) (2009).

26  Gao, W. & Huang, R. Thermomechanics of monolayer graphene: Rippling, thermal expansion and elasticity. *J. Mech. Phys. Solids* **66**, 42-58 (2014).

27  Neek-Amal, M., Abedpour, N., Rasuli, S.N., Naji, A. & Ejtehadi, M.R. Diffusive motion of C-60 on a graphene sheet. *Phys. Rev. E* **82**, 051605 (2010).

28  Lebedeva, I.V. *et al.* Fast diffusion of a graphene flake on a graphene layer. *Phys. Rev. B* **82**, 155460 (2010).

29  Los, J.H., Bichara, C. & Pellenq, R.J.M. Tight binding within the fourth moment approximation: Efficient implementation and application to liquid Ni droplet diffusion on graphene. *Phys. Rev. B* **84**, 085455 (2011).

30  Maruyama, Y. Temperature dependence of Levy-type stick-slip diffusion of a gold nanocluster on graphite. *Phys. Rev. B* **69**, 245408 (2004).

31  Lewis, L.J., Jensen, P., Combe, N. & Barrat, J.L. Diffusion of gold nanoclusters on graphite. *Phys. Rev. B* **61**, 16084-16090 (2000).

32  Chen, J. & Chan, K.Y. Size-dependent mobility of platinum cluster on a graphite surface. *Mol. Simul.* **31**, 527-533 (2005).

33  Bao, W. *et al.* Controlled ripple texturing of suspended graphene and ultrathin graphite membranes. *Nature Nanotech.* **4**, 562-566 (2009).

34  Xu, P. *et al.* Unusual ultra-low-frequency fluctuations in freestanding graphene. *Nat. Commun.* **5**, 3720 (2014).

35  Supplementary information.

36  Bonini, N., Garg, J. & Marzari, N. Acoustic Phonon Lifetimes and Thermal Transport in Free-Standing and Strained Graphene. *Nano Lett.* **12**, 2673-2678 (2012).

37  Meyer, J.C. *et al.* The structure of suspended graphene sheets. *Nature* **446**, 60-63 (2007).

38  Zan, R. *et al.* Scanning tunnelling microscopy of suspended graphene. *Nanoscale* **4**, 3065-3068 (2012).

39  Bangert, U. *et al.* STEM plasmon spectroscopy of free standing graphene. *Phys. Status Solidi A* **205**, 2265-2269 (2008).

40  de Lima, A.L. *et al.* Soliton instability and fold formation in laterally compressed graphene. *Nanotechnology* **26**, 045707 (2015).

41  Guo, Y. & Guo, W. Soliton-like thermophoresis of graphene wrinkles. *Nanoscale* **5**, 318-323 (2013).

42  Cao, P.G., Xu, K., Varghese, J.O. & Heath, J.R. The Microscopic Structure of Adsorbed Water on Hydrophobic Surfaces under Ambient Conditions. *Nano*

*Lett.* **11**, 5581-5586 (2011).

43  Tocci, G., Joly, L. & Michaelides, A. Friction of water on graphene and hexagonal boron nitride from ab initio methods: very different slippage despite very similar interface structures. *Nano Lett.* **14**, 6872-6877 (2014).

44  Zambrano, H.A., Walther, J.H., Koumoutsakos, P. & Sbalzarini, I.F. Thermophoretic motion of water nanodroplets confined inside carbon nanotubes. *Nano Lett.* **9**, 66-71 (2009).

45  Abascal, J.L.F. & Vega, C. A general purpose model for the condensed phases of water: TIP4P/2005. *J. Chem. Phys.* **123**, 234505 (2005).

46  Jiang, T., Huang, R. & Zhu, Y. Interfacial Sliding and Buckling of Monolayer Graphene on a Stretchable Substrate. *Adv. Funct. Mater.* **24**, 396-402 (2014).

47  Yang, L.Q. & Rahman, T.S. Enhanced anharmonicity on Cu(110). *Phys. Rev. Lett.* **67**, 2327-2330 (1991).

48  Ciplys, D. *et al.* in *2010 IEEE Sensors* 785-788 (2010).

49  Mayo, S.L., Olafson, B.D. & Goddard III, W.A. Dreiding - a generic force-field for molecular simulations. *J. Phys. Chem.* **94**, 8897-8909 (1990).

50  Plimpton, S. Fast parallel algorithms for short-range molecular-dynamics. *J. Comp. Phys.* **117**, 1-19 (1995).

**Methods**

Full details of the computational set-up are given in Sections 1 in the SI[35] and only the key features are reported here. All results reported in the main manuscript have been obtained from force field based MD simulations. Although we have also performed *ab initio* MD simulations, which are discussed in Section 14 in the SI, force fields are the method of choice because of the large system sizes (thousands of atoms) and the long timescale dynamical simulations. Results are reported for graphene simulation cells comprised of 8,540 carbon atoms (15×15 nm). Additional results with larger water droplets (up to 14 nm diameter, with almost 32,000 molecules) and different sized unit cells in which the sensitivity of the results to finite size effects was examined are also provided in Sections 3 and 12 in the SI[35].

We use the TIP4P/2005[45] potential for water and the Dreiding force field[49] for graphene. The van der Waals interaction between water and graphene is described by a Lennard-Jones (L-J) 6-12 potential between the oxygen and carbon atoms, the parameters of which are based on the interaction energy of a water monomer on graphene obtained from a recent diffusion quantum Monte Carlo study[16]. With our particular set-up this yields water-graphene contact angles for the nanodroplets in the 90-93° range. Upon extrapolation of the size dependence of the contact angle to macroscopic droplets we obtain a contact angle of 89°±0.5° (see Section 6 in the SI[35]). The predicted contact angle, which does not involve a fit to experiment as is normally done in MD simulations of water on graphene, lies within the 74° to 127° range of experimental values reported[10]. The $C_{60}$-graphene interaction is also described with a L-J 6-12 potential with the parameters taken from Ref.[27].

All systems were equilibrated in the NVT ensemble at 298 K for 2 ns by applying separate Nosé-Hoover chain thermostats to water and graphene. Following this, the thermostat on the water molecules was removed, and the systems were evolved for at least 35 ns and in several circumstances much longer. Such long runs are required to obtain accurate diffusion coefficients (see Section 8 in the SI[35]). A careful series of tests established that this thermostating set-up was the most appropriate simulation protocol for the class of system under investigation, as discussed in Section 4 in the SI[35]. Following Gao and Huang's method[26], the lattice constant of the simulation cell used for the small amplitude ($A$ = 0.2 nm) ripples was obtained from an initial energy minimization at 0 K (corresponding to a zero strain state). All MD simulations reported were carried out with the LAMMPS code[50].

As well as the tests mentioned already, extensive tests of the computational set-up have been performed including comparison to experiments whenever possible, stability of the droplets with respect to evaporation, validation of the force fields and results with other force fields, and sensitivity of the results to the presence of defects. These are all discussed in the SI[35].